\documentclass[aps,apl,preprint,floatfix,amsmath,amssymb,showpacs,showkeys]{revtex4}
 
\newcommand{\be}{\begin{equation}}
\newcommand{\ee}{\end{equation}}
\newcommand{\ba}{\begin{align}}
\newcommand{\ea}{\end{align}}
\newcommand{\bn}{\begin{eqnarray}}
\newcommand{\en}{\end{eqnarray}}

\def\ba{\begin{eqnarray}}
\def\ea{\end{eqnarray}}

\def\({\left(}
\def\){\right)}
\def\g[{\left[}
\def\g]{\right]}
\def\g{{\left{}
\def\g}{\right}}

\usepackage{graphicx}

\begin{document}


\title{Klein tunneling of spin polarized currents in HgTe quantum wells}

\author{D. R. Viana}
\email{davidson.viana@ufv.br}

\affiliation{Faculdade de Vi\c{c}osa, 36570-000, Vi\c{c}osa, Minas Gerais, Brazil.}

\author{T. M. Melo}
\affiliation{Instituto Federal do Rio de Janeiro, 27541-030, Resende, Rio de Janeiro, Brazil.}

\author{J. M. Fonseca}
\email{jakson.fonseca@ufv.br}
\affiliation{Departamento de F\'{i}sica, Universidade Federal de Vi\c{c}osa, 36570-000, Vi\c{c}osa, Minas Gerais, Brazil.}

\author{D. S. Souza}
\affiliation{Instituto Federal Farroupilha, 97110-767, Santa Maria, Rio Grande do Sul, Brazil.}

\author{W. A. Moura-Melo}
\email{winder@ufv.br}
\homepage{https://sites.google.com/site/wamouramelo/}
\affiliation{Departamento de F\'{i}sica, Universidade Federal de Vi\c{c}osa, 36570-000, Vi\c{c}osa, Minas Gerais, Brazil.}

\author{A. R. Pereira}
\email{apereira@ufv.br.}
\homepage{https://sites.google.com/site/quantumafra/home}
\affiliation{Departamento de F\'{i}sica, Universidade Federal de Vi\c{c}osa, 36570-000, Vi\c{c}osa, Minas Gerais, Brazil.}

\date{\today}
\begin{abstract}
We investigate the behavior of spin polarized currents in two-dimensional topological insulators (TI). Stationary solutions inside a HgTe/CdTe quantum well (QW) were  obtained by Bernevig-Hughes-Zhang (BHZ) model modified by a electric and magnetic barrier inside a non-completely insulating bulk. An attenuated quantum spin Hall (QSH) effect occurs in the gaped region with an apparent Klein-like paradox. Even more interesting, for strong potential regime, the interaction between the quasiparticles and the barriers allows spin inversion of this electronic states in a distinct channel conduction. Thus, our findings suggest a mechanism to manipulated spin polarized currents in this system.
\end{abstract}
\maketitle

\section{Introduction and Motivation}

Recently, a new class of topological state of matter has emerged, named topological insulators or QSH insulators \cite{moore, review, left}.  They have been predicted to exhibit exotic physical properties depending only on their underlying topology and not on its particular geometry nor mechanical features. QSH insulators promising candidates for spintronic technologies \cite{livro,giant,design,moore} and optical applications \cite{PLA, otica} due to strong correlation between spin and momentum (large spin-orbit coupling) in these compounds . The two-dimensional (2D) TI exhibit a peculiar metallic edge firstly discovered in HgTe/CdTe QW \cite{konig2007}. This QSH insulators are characterized by a full insulating gap in the bulk while their edge states present metallic, chiral, practically dissipationless and gapless Dirac dispersion similar to graphene. Bernevig, Hughes and Zhang developed an effective Hamiltonian, so called BHZ model \cite{BHZ2006}, for describing these special states and they also predicted a quantum phase transition in QW as a function of its width (with a critical value), rendering a single pair of helical edge states.
%

Here, we shall consider BHZ model augmented by a step-like barriers, which may mimic a gate-potential or a magnetic film to study its effects on the 2D-TI electronic states in a non-completely insulator bulk. We propose a particular situation with attenuated quasiparticle currents flowing through the edge and toward the bulk where the barriers is inserted, blocking and flipping partially spin polarizations. The fermions that return with inverted spin travel in the opposite direction from the incident current feeding a different channel, where the Dirac mass term is zero \cite{review,left}. The amplitude of this effect can be tuned by the potential barriers. Thus, our findings suggest a mechanism to manipulated spin polarized currents in HgTe/CdTe QW.

\section{The Model and its Basic Features}

The BHZ model is designed to describe the behavior of electronic states in a HgTe/CdTe heterostructure QW having a width $d$. The effective Hamiltonian describing the stationary states inside QW created by the energy difference between the nearest sub-bands of the Fermi level in a inverted regime \cite{review}, reads like below:

\begin{equation}\label{H}
{H}_{\rm eff}(k) = 
\left[
\begin{array}{cccc}
 {\cal H}(k) & 0 \\
0 & {\cal H}^*(-k) \\
\end{array}
\right],
\end{equation}\\
where ${\cal H}(k)=\epsilon(k){\bf 1}_{2\times2} + \sum_i \xi_i(k)\sigma_i $, $\epsilon(k)=C-D(k_x^2 +k_y^2)$, ${\bf 1}_{2\times2}$ and $\sigma_i$ are the identity and the Pauli matrices, respectively. $\xi_1=Ak_x $, $\xi_2=Ak_y$, are the hybridization terms, $\xi_3=M-B(k_x^2 +k_y^2)$ is a mass-like term where $B$ and $M$ are known as Newtonian and Dirac mass-like parameters (more details, below). $A$, $B$, $C$ and $D$ are tunable experimental parameters that depend on the quantum well geometry. $C$ is the minimum energy or Fermi energy near to $\Gamma$-point (we normalize energy by taking $C=0$). We also set $D=0$ once it has no effect on phase transition and on topological properties; in addition, its vanishing ensures particle-hole symmetry \cite{review}.  $M$ equals the difference between two energies nearest the Fermi level, $|E_1,+(-)\rangle $ and $|H_1,+(-)\rangle $. If this gap vanishes, what happens at the so-called $\Gamma$ point, one has two copies of massless Dirac Hamiltonian, with doubly degenerate states, each of them accounting for a given spin polarization. In the absence of the external fields and impurities, $M$ is positive provided that HgTe QW width, $d$ is smaller than a critical value, $d_c \approx 6.3 \,nm$. In this case, an insulating phase occurs. Whenever $d_c$ threshold is overcome one gets $M<0$ yielding a conducting phase whose states move on the edges exhibiting spin Hall effect \cite{konig2007,konig2008,review}. 

Additional terms concerning other features could be considered in Eq. (\ref{H}), for instance, bulk inversion asymmetry (BIA), which predicts the phenomenon known as band inversion and structural inversion asymmetry (SIA), related to boundary conditions of asymmetrical quantum well. The first is small compared to the hybridization terms, (see Eq.(\ref{H})). While SIA-term is minimized when we have a symmetric quantum well \cite{dre1, type, review}, however we do not take it account in our approuch by simplicity.

\section{Quantum Wells of $HgTe/CdTe$ with a electrostatic potential barrier}

We consider an electrostatic barrier largely distributed in HgTe QW width, assuming the band inversion in this region, i. e. $d>d_c$. Such potential barrier can be an applied back gate voltage or a large electrostatic impurity. To study the dynamics along $x$ we introduce a step-like potential barrier   according Fig. \ref{fig1}. Once translational symmetry is preserved along $y$ (free motion direction, no barrier), we take $k_y = 0$ in Eq. (\ref{H}), so that our proposal may be explicitly written as:
 
\begin{eqnarray}\label{H+V}
& & H_{\rm bulk}= H_{0}(k_x) + {\bf 1}_{4\times 4} V(x)= \nonumber\\ 
\nonumber\\
& & \hspace{-0.7cm} \left[ \begin{array}{cccc}
 \tilde{M}+V(x) & Ak_x & 0 & 0 \\
  Ak_x & -\tilde{M}+V(x) & 0 & 0\\
0 & 0 & \tilde{M}+V(x) & -Ak_x \\
0 & 0 & -Ak_x & -\tilde{M}+V(x) \\
\end{array} \right],\, \, \,
\end{eqnarray}
\noindent with $\tilde{M}(k_x)=M-Bk^2_x$, while $V(x)=V_0 \Theta(x-x_0)$ is intended to manipulate the energy of the Dirac fermions near the edge (similarly to chemical potential modifying the Fermi level) and to create a mechanism to control the attenuation of the metallic edge states. 

\begin{figure}[h]\label{fig1} \centering
\includegraphics[height=7cm]{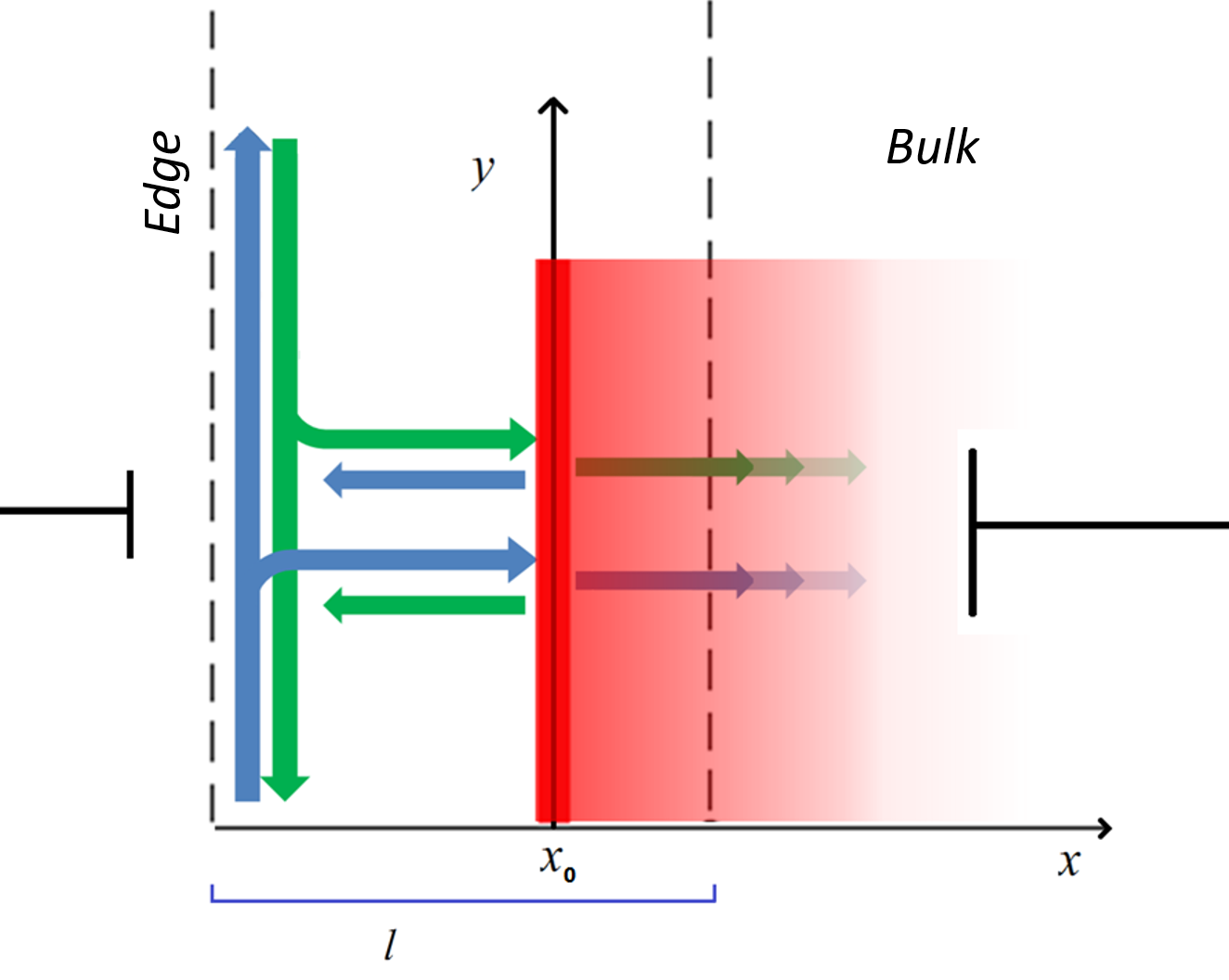}
\caption{\small Illustration of the spin up (green) and spin down (blue) currents leaving and returning from the edge. The red rectangle represents the electric potential barrier produced by a back gate voltage, described by $V(x)=V_0\Theta(x-x_0)$. The indicated $l$ represents the decay length of edge states without barrier influence. Part of reflected currents have reverted spins and feed the zero-gap currents in the the edge again, increasing the conductivity in a different channel.}
\label{fig1}
\end{figure}


In order to obtain the dispersion relation of the edge states, we write the eigenvalue equation: 

\begin{equation}\label{Heff}
H_{\rm bulk} \big(k_x \rightarrow i\partial_x \big) \Psi(x) = (E-V(x)) \Psi(x)\, ,
\end{equation}\\
where $E$ is the energy eigenvalues of the stationary solutions for incident particles in the absence of electrostatic barrier. The eingenstates of Eq. (\ref{Heff}) describe opposite spins related by time reversal symmetry, $\mathcal{T} \Psi_{\uparrow}(x)=\Psi_{\downarrow}(x)$. We introduce the $ansatz$, $\Psi(x) \propto e^{k x}$, according the Ref. \cite{review}. We choose a incident wave incoming from the right side to the left so that the wavenumber is the coefficient $k$, related with the experimental parameters as $k=\frac{1}{2B}(A + i\sqrt{4MB -A^2})$. We propose a theoretical situation where $4MB >A^2$, enabling an attenuated and oscillatory solution in the bulk. Such regime is obtained by follow the tendency values of the experimental data associated to the larger thickness of the QW \cite{review}. For the two different regions in the bulk, say, outside ($x < x_0$) and inside the barrier, $(x\ge x_0)$ we obtain the incident ({\em i}), reflected ({\em r}) and transmitted ({\em t}) solutions, as follows:
 
\begin{equation}\label{psiinc}
\Psi_i  = {\cal A} ^{\uparrow}_i e^{kx} \left[
\begin{array}{cccc}
 1 & \\
\frac{Ak_1}{E+\tilde{M}(k_1)} &\\
0 & \\
0 &
\end{array} \right],\,
\end{equation}

\begin{equation}\label{psiref}
\Psi_r  = {\cal A} ^{\uparrow}_r e^{-k_r(x-x_0)} \left[
\begin{array}{cccc}
 1 & \\
\frac{-Ak_1}{E+\tilde{M}(k_1)} &\\
0 & \\
0 &
\end{array} \right] \,
+
{\cal A}^{\downarrow}_r e^{-k_r(x-x_0)} \left[
\begin{array}{cccc}
0 &\\
0 &\\
 1 & \\
\frac{-Ak_1}{E+\tilde{M}(k_1)} &
\end{array} \right],\,
\end{equation}

\begin{equation}\label{psitrans}
\Psi_t  = {\cal A}^{\uparrow}_t e^{k'(x-x_0)} \left[
\begin{array}{cccc}
1 & \\
\frac{Ak_2}{E-V_0 +\tilde{M}(k_2)} & \\
0 & \\
0 &
\end{array}
\right] \,
+
{\cal A}^{\downarrow}_t e^{k'(x-x_0)} \left[
\begin{array}{cccc}
0 & \\
0 & \\
1 & \\
\frac{Ak_2}{E-V_0 +\tilde{M}(k_2)} & 
\end{array}
\right],
\end{equation}\\
where $k=a+ik_1$, $k_r=a_r+ik_1$ and $k'=a'+ik_2$ are the modules of the incident, reflected and transmitted wavenumbers, respectively. Each stationary solution corresponds to currents of particles with momentum $\hbar k_x\equiv \hbar k$. The BHZ model describes spin up in conduction band and spin down particles in valence band. Once the edge states has a decay length ($l$), the bulk can not be a complete insulator, as seen in the Ref. \cite{now2013}. Therefore, the behavior of the spin polarized currents can be manipulated by mean of the proposed barriers (electrostatic and magnetostatic). Since the constants $ a <0 $, $ a '<0 $ and $ a_r> 0 $, the attenuation lengths are $ l_1 = 1/| a|$, $l_r=1/ |a_r | $ and $l_2 = 1/ | a'|$. Without the existence of the electrostatic barrier, the value of $a$ is equal a $ A/2B$. 


The complete combined eigensolution for Eq. \ref{Heff} is:

\begin{equation}\label{complete}
\Psi(x) =\Theta(x+x_0)\left[\Psi_i ^{\uparrow}(x) +\Psi_r ^{\uparrow}(x) + \Psi_r ^{\downarrow}(x)\right] +\Theta(x-x_0)\left[\Psi_t ^{\uparrow}(x) +\Psi_t ^{\downarrow}(x)\right].
\end{equation}\\
This solution inserted in the Eq. \ref{Heff} leads to the dispersion relation (for both spins up and down) below:
\begin{equation}
\left\{ \begin{array}{l}   
 E^2 = A^2k_1^2 +\tilde{M}^2(k_1) \, \, \, \, \, \, \, \, \, \, \, \,  \, \, \, \, \, \, \, \, \, \, \,   {\mbox{\rm  (outside barrier)}} \\
\(E-V_0\)^2= A^2k_2^2+\tilde{M}^2(k_2) \, \, \,\, {\mbox{\rm (inside barrier)}}. 
\end{array}     \right.
\end{equation}\\
By mean of boundary conditions imposed in $\Psi(x=x_0)$ (Eq. \ref{complete}), we obtain relations on the normalization factors: $\mathcal{A}_i ^{\uparrow}=(1+\Omega) \mathcal{A}_t^{\uparrow}$,  $\, 2\mathcal{A}_r^{\uparrow}=(1-\Omega)\mathcal{A}_t^{\uparrow}$ e $\, \mathcal{A}_r^{\downarrow}=-\Omega\mathcal{A}_t^{\downarrow}$, where $\Omega= \frac{k'}{k} \frac{E+\tilde{M}(k_1)}{E+\tilde{M}(k_2)-V_0}$. As seen, the solution admits nonzero amplitudes of inverted spins. This situation is analogous to Dresselhaus effect for a semiconductor bulk, whose scattered currents depends on the spin polarization due to spin-orbit interaction \cite{dre1,dre2,dre3}. Here, we consider a potential  regime strong enough to invert the spin redirecting the particles to a different channel for part of the reflected spin polarized currents (see Fig. \ref{fig1}).

The dispersion and the eigensolutions leads to the associated average spin polarized (density) currents may be calculated by the standard quantum mechanical procedure, $\langle \hat{J_x}\rangle=\Psi^{\dagger } \frac{\partial H_{bulk}(k_x)}{\partial k_x} \Psi$, and for both spins, read explicitly:

\begin{eqnarray}\label{Ji}
& & J_{i}^{\uparrow}= 2|{\cal A}_{i}^{ \, \uparrow}|^2 \left\{ Bk_1 +\frac{A^2 k_1 }{E+\tilde{M}(k_1)} +\frac{BA^2 k_1^3 }{\left[E+\tilde{M}(k_1)\right]^2} \right\} e^{2|a|(x-x_0)}  \nonumber \\
& & (x< x_0),
\end{eqnarray}
\begin{eqnarray}\label{Jr}
& & J_{r}^{\uparrow, \, \downarrow}=2|{\cal A}_{r}^{ \, \uparrow, \, \downarrow}|^2  \left\{ Bk_1 +\frac{A^2 k_1}{E+\tilde{M}(k_1)} +\frac{BA^2k_1^3 }{\left[E+\tilde{M}(k_1)\right]^2} \right\} e^{2|a_r|(x-x_0)} \nonumber \\
&  & (x < x_0),
\end{eqnarray}
\begin{eqnarray}\label{Jt}
& & J_{t}^{\uparrow, \, \downarrow}=2|{\cal A}_{t}^{ \, \uparrow, \, \downarrow}|^2 \left\{ Bk_2 +\frac{A^2 k_2}{E-V_0+\tilde{M}(k_2)} +\frac{BA^2k_2^3}{\left[E-V_0+\tilde{M}(k_2)\right]^2} \right\}e^{-2|a'|(x-x_0)}, \nonumber \\
& & (x\geq x_0)\,.
\end{eqnarray}\\
Is imposed that total current in the frontier  ($x=x_0$)  is conserved: $J_{i \, \uparrow}= J_{r \, \uparrow}+ J_{r \, \downarrow}+ J_{t \, \uparrow}+J_{t \, \downarrow}$. Analogously, the (average) non-gaped  edge currents $\langle J_{edge}\rangle$ can be obtained by edge hamiltonian, as follow: 

\begin{equation}
 J_{y}^{ \pm, \uparrow \, \downarrow}=\Psi^{\dagger } \frac{\partial H_{edge}(k_y)}{\partial k_y} \Psi = \pm 2Bk_y|{\cal A}_y|^2 \left[ 1 +\frac{A^2k_y^2 }{{\left(E-Bk_y^2\right)^2}} \right].\\
\end{equation}\\
From of the currents we obtain the spin up reflectance and transmittance:
\begin{equation}\label{ref}
R^\uparrow=\frac{|1-\Omega|^2}{|1+\Omega|^2}e^{-2(|a|-|a_r|)(x-x_0)}, \,\,\, (x< x_0),
\end{equation}
\begin{equation}\label{trs}
T^\uparrow=4 \frac{k_2}{k_1}\frac{E+\tilde{M}(k_1)}{E+\tilde{M}(k_2)-V_0} \frac{e^{-2(|a|-|a'|)(x-x_0)}}{|1+ \Omega |^2}, \,\,\,(x\geq x_0).
\end{equation}\\
A weak ($V_0<E+\tilde{M}(k_2)$), or a intermediary potential ($E-\tilde{M}(k_2)< V_0 < E+\tilde{M}(k_2)$), do not change the spin polarization of scattered currents. To support this, in terms of probability conservation, for spin up currents we have $R^\uparrow+T^\uparrow=1$. 

The electrostatic potential $V_0$ can be manipulated to obtain three distinct scenarios depending on the relative value of the `net energy', $E+\tilde{M}(k_2)-V_0$. A weak potential leads small attenuation, once assumed $a\sim a'$ (if $V_0 \ll E$, we have $l_t \approx |2B/A|$). With an intermediary value of $V_0$ the wavenumber inside the barrier ($k_2$) becomes pure imaginary, modifying the transmitted current and hence the transmittance. Thus, the decay length of edge states inside the barrier is $l_{t}=\frac{1}{|a'-k_2|}$. In this particular case, $ k '=a'-k_2 $ is a real number and the wave inside the barrier is completely attenuated. This situation is analogous to the insulating bulk where $A^2>4MB$ in the inverted regime \cite{review}.

A strong potential, $V_0>E + \tilde{M}(k_1)$, yields to a peculiar oscillatory regime regarded (with $A^2<4MB$). There are some consequences of a non completely insulating or/and a semiconductor bulk. The most relevant is the possibility of negative transmittance due to the sign of the $\Omega$ in the Eq. (\ref{trs}). The interpretation is analogous to a Klein paradox which a intense potential leads to a particles propagating in opposite direction (consequently, the wavenumber $k_2$ becomes negative). Differently of this usual case, there are a attenuation in this system depending on experimental parameters $A$ and $B$, once we assume $a\neq 0$ and $a' \neq 0$. In addiction, the geometry of the QW and the potential regime determine the dynamic of the spin currents associated to collective excitations according the BHZ model parameters. These particular gapped quasiparticles propagate as Dirac fermions and holes towards the bulk, similar to particles and antiparticles in ordinary case. Klein tunnelling was also investigated in 3D TI as reported in Ref. \cite{spinsig}.

\subsection{Velocities}

\begin{figure}[h]
 \centering
\includegraphics[height=7cm]{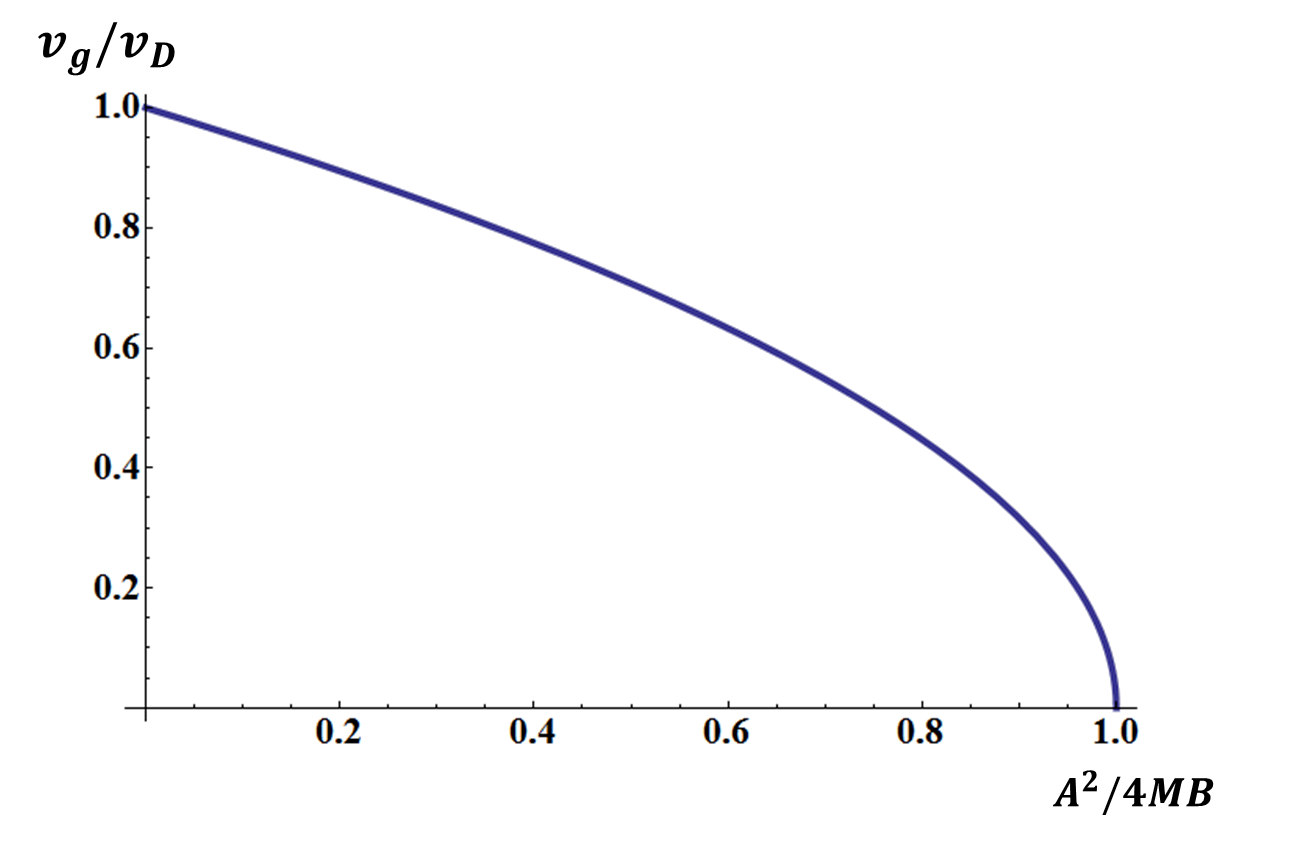}
\caption{\small This graphic is the ratio between group velocity and Dirac velocity as a function of the ratio of the geometrical parameters.}
\label{g1}
\end{figure}

When propagating in the bulk, the spin polarized currents carries geometric characteristics of QW. So the group $\big(\frac{1}{\hbar}\frac{dE}{dk_2}\big)$ and the phase ($E/\hbar k$) velocities have a dependence on the geometric parameters modifying the velocities of edge and bulk states. The group velocities outside and inside the barrier are respectively:
\begin{equation}
|v_{g1}| = \frac{|A|}{\hbar}\sqrt{ 1-A^2/4MB} \leqslant v_{D}= |A|/\hbar,
\end{equation}
\begin{equation}
|v_{g2}| = \frac{1}{\hbar}\frac{A^2-2MB+2Bk_2^2}{\sqrt{A^2-2MB+Bk_2^2+M^2/k_2^2}},
\end{equation}\\
where $v_D$ is a constant along the edge, known as  Dirac velocity (for a QW with a width of $7 \, nm$, $v_D \approx 5.5 \times 10^7 \, cm/s$). This is the maximum value of the velocity of the quasiparticles in the inverted regime in BHZ model. The behavior of group velocity outside the barrier is illustrated in the Fig. \ref{g1}. The group velocity inside the barrier can be modulated by $V_0$ once $k_2$ depends on it. The product  $4MB$ increases with the width and $A^2$  have small variations. Consequently, the model predicts that in a QW with infinity width, the velocity in the bulk would be equal to Dirac velocity of the edge states \cite{review}. In this case we would have a system that behaves like a infinity homogeneous isotropic nondispersive semiconductor plane. In the absence of electrostatic barrier $v_{g2}$ reduces to $v_{g1}$ whose threshold case is $v_{g1}=v_D$.

Phase velocities outside ($v_{p\, 1}$) and inside  ($v_{p\, 1}$) the electrostatic barrier are subsequently:

\begin{equation}
 v_{p\, 1}=\pm \sqrt{v_D^{\,2}-\frac{\tilde{M}(k_1)^2}{\hbar^2 k_1^{\, 2}}},
\end{equation}

\begin{equation}
v_{p\, 2}=\pm \sqrt{\(v_D-\frac{V_0}{\hbar^2 k_2^2}\)^2-\frac{\tilde{M}(k_2)^2}{\hbar^2 k_2^{\, 2}}}.
\end{equation}\\
Similarly to electrons in a periodic potential, the bulk states phase velocity also has a maximum value, $v_D$, outside barrier. Above this limit, the topological regime reached in the HgTe QW is extrapolated to a nondispersive semiconductor and the BHZ model is not applicable. In addiction, inside the barrier, the velocity can be manipulated by the scalar potential $V_0$ and the upper limit depends on it. Peculiarly, the phase velocity inside and outside the barrier not depends directly of geometrical parameter $A$. 

\section{Inclusion of a Magnetostatic Barrier}

We inserted a magnetostatic barrier in the same region of the included electrostatic step potential case. This situation can be created by a magnetic impurity largely distributed along the region $x>x_0$ enabling a interaction with the conduction quasiparticles. This is performed by Landau gauge \cite{sakurai} that reads  $\vec{A}_L=(0, A_y, 0)$, where $A_y(x)=\Theta(x-x_0)B_z x$, and $B_z$ is a constant  magnetic field perpendicular to the plane of heterostructure of HgTe (plane of conduction). The eigenvalue equation with this prescription, becomes:

\begin{equation}\label{eqHBz}
H_{\rm bulk} \bigg(k_x \rightarrow i\partial_x -\frac{q}{\hbar c}A_y (x)\bigg) \Psi(x) = (E-V(x)) \Psi(x).\,
\end{equation}\\
This approach implies a time reversal symmetry breaking, leading to spinor eingensolutions dependent of the magnetic field and the scalar potential. The hamiltonian is also separable in part independent of $k_y$ such as the previous case. We obtain analogous solutions of the step scalar potential. The most significant changes are the wavenumber  $k_2$, which depends of $B_z$ and the mass term becomes: $\tilde{M}(k_2)=M-B(k_2 -\frac{q}{\hbar c}A_y)^2$. The currents (Eqs. (\ref{Ji}), (\ref{Jr}), (\ref{Jt}) and the coefficients (\ref{ref}) and (\ref{trs}) ) follow these changes reducing the effective wavenumber inside the magnetostatic barrier. 


\section{Conclusions and Final Remarks}

In summary, the modified BHZ model with insertion of electrostatic step potential in HgTe QW yields to distinct energy regimes. For a weak potential we have oscillatory solutions corresponding to a peculiar situation where spin currents are a little attenuated, similar to a semiconductor bulk. Considering an intermediary potential regime the quasiparticles beam is completely attenuated resembling a topological insulator described by a non-modified BHZ model in a QW with $A^2>4MB$. In both cases there is no spin reversion inside and outside the barrier. 

When the scalar potential its strong enough (considering $A^2<4MB$), such regime has the interesting possibility of spin reversion of the quasiparticles  that propagates towards the bulk. This effect is described by an apparent negative transmittance analogous to a Klein paradox. The dynamics of attenuated spin currents depending on the geometry of QW. In this context the intensity of beam scattered can be manipulated by barrier potential.

The  group and phase velocities of spin currents depends on the geometric parameters of QW. Outside the barrier, the group velocity is limited by $v_D$:  when width of QW tends to infinity, the whole system behaves like a homogeneous nondispersive semiconductor plane. Still outside the barrier, we also have a threshold value given by $v_D$ for phase velocity. Inside the step potential, the group and phase velocities can be indirectly manipulated by $V_0$, once both velocities depends on momentum that is related to $V_0$ through dispersion relation.

For magnetostatic barrier, the relevant modification is the dependence on Bz field of wavenumber inside the barrier and the mass term $\tilde{M}(k_2)$. Therefore, spin polarized currents, transmittance and reflectance of the quasiparticles can be manipulated by intensity of magnetostatic barrier. The applied magnetic field   reduces the wavenumber by mean of the Landau gauge inside this barrier diminishing the tunneling.

The control of electrotastic and magnetostatic barrier can improve spintronic applications, once our system enable to manipulate the spin information carried by quasiparticles. Our proposal described polarized spin currents via BHZ model. Some experimental detections of QSH effect in HgTe/CdTe in Refs. \cite{left, Spinnature, brune2010} are examples of possibilities of manipulation of spin currents with opposite direction in these systems. In addiction, similar detections can be find in other materials \cite{kato2004, review, livro}. 

\section{Acknowledgments}
The authors thank CAPES, CNPq and FAPEMIG (Brazilian agencies) for partial financial support.

\end{document}